\documentclass[aip, print]{revtex4-1}

\usepackage{graphicx}
\usepackage{dcolumn}
\usepackage{bm}
\usepackage{sidecap}
\usepackage{xcolor}
\draft 

\begin{document}

\title{Hyper Resolute Ultra thin Low Cost All-Dielectric Broadband Achromatic Metalenses} 

\author{Giuseppe Emanuele Lio}
\affiliation{CNR Nanotec-Institute of Nanotechnology, UOS Cosenza, 87036 Rende (CS), 87036, Italy}
\affiliation{Physics Department, University of Calabria, I-87036 Arcavacata di Rende (CS), Italy}
\author{Antonio Ferraro}
\email[]{E-mail: antonio.ferraro@unical.it}
\affiliation{Physics Department, University of Calabria, I-87036 Arcavacata di Rende (CS), Italy}
\affiliation{CNR Nanotec-Institute of Nanotechnology, UOS Cosenza, 87036 Rende (CS), 87036, Italy}
\author{Tiziana Ritacco}
\affiliation{CNR Nanotec-Institute of Nanotechnology, UOS Cosenza, 87036 Rende (CS), 87036, Italy}
\affiliation{Physics Department, University of Calabria, I-87036 Arcavacata di Rende (CS), Italy}
\author{Dante Maria Aceti}
\affiliation{Physics Department, University of Calabria, I-87036 Arcavacata di Rende (CS), Italy}
\author{Antonio De Luca}
\affiliation{Physics Department, University of Calabria, I-87036 Arcavacata di Rende (CS), Italy}
\affiliation{CNR Nanotec-Institute of Nanotechnology, UOS Cosenza, 87036 Rende (CS), 87036, Italy}
\author{Roberto Caputo}
\email[]{E-mail: roberto.caputo@unical.it}
\affiliation{Physics Department, University of Calabria, I-87036 Arcavacata di Rende (CS), Italy}
\affiliation{CNR Nanotec-Institute of Nanotechnology, UOS Cosenza, 87036 Rende (CS), 87036, Italy}
\affiliation{Institute of Fundamental and Frontier Sciences, University of Electronic Science and Technology of China, Chengdu 610054}
\author{Michele Giocondo}
\affiliation{CNR Nanotec-Institute of Nanotechnology, UOS Cosenza, 87036 Rende (CS), 87036, Italy}

\begin{abstract}
Metalenses offer the ground-breaking opportunity to realize highly performing low-weight, flat and ultrathin, optical elements which substantially reduce size and complexity of imaging systems. Today, a major challenge in metalenses design is still the realization of achromatic optical elements ideally focussing a broad wavelength spectrum at a single focal length.  Here we present a fast and effective way to design and fabricate extremely thin all-dielectric metalenses, optimally solving achromaticity issues by means of  machine learning codes. The enabling technology for fabrication is a recently developed hyper resolute two-photon direct laser writing lithography equipment. The fabricated metalenses, based on a completely flat and ultrathin design, show intriguing optical features. Overall, achromatic behavior, focal length of 1.14 mm, depth of focus of hundreds of microns and thickness of only few nanometers allow considering the design of novel and efficient imaging systems in a completely new perspective.
\end{abstract}


\maketitle 
\section{Introduction}
The development of optical elements with nanoscale features has recently raised a very huge and wide interest in materials science \cite{genevet2017recent}. 
These research efforts are motivated by fundamental physical insights demonstrating the possibility to realize perfectly flat lenses with eventual application in a plethora of optical systems including integral imaging, phase filters, discrete lenses for sensing, high numerical aperture and achromatic optical devices, optical tunability or reconfigurability, phase dispersion correction and correlation and augmented-reality (AR) applications \cite{chen2018broadband, chen2019broadband, fan2019broadband, idjadi2020nanophotonic, gao2013plasmonic, yu2014flat, byrnes2016, khora2017nano, khorasaninejad2017metalenses, cui2019tunable, chen2020flat}. 
Initially, studies considered metasurfaces where linear gradients of phase discontinuities lead to planar reflected and refracted wavefronts. These metasurfaces, mainly working in the IR spectrum, were characterized by discrete patterns involved to mime the functionality of classical refractive lenses thus reconstructing the phase shift typically introduced by the surface curvature \cite{khora2015achromatic, khora2016science, khora2017science}.
The replacement of bulky refractive optics with diffractive planar elements represents an important leap forward for the miniaturization of optical systems. However, by their nature, diffractive optics suffer of chromatic aberrations due to the phase dispersion ($\theta$) accumulated by  light during its propagation within the optical element. It has been largely demonstrated that this limitation can be overcome when a suitable wavelength-dependent phase shift is directly imparted to the metasurface distribution. 
However, this option requires considering focal length ($f$) and beam deflection under a new point of view. In fact, period ($\Lambda$), number of the discrete elements ($N$) involved in the final structure and radius of the area occupied by the metalenses ($R_{N}$) become critical specifications \cite{aieta2015multiwavelength}. Similar metalenses, realized by Capasso et al., were initially constituted by metal sub-elements replaced later on by dielectric ones: an example is given by an array of $SiO_2$ nanoscopic conical bulges with base diameter of few hundreds of nanometers and height of around $2\mu m$\cite{park2019all}. While showing noticeable optical features, these metalenses require advanced lithography techniques for their fabrication that makes manufacturing on a large area extremely slow and time-consuming \cite{she2018large}. In the last years, much effort has been devoted to improve metalenses and metasurfaces for several applications, thus involving them as diffractive elements \cite{ni2013ultra, arbabi2015dielectric, arbabi2015subwavelength, naserpour2017recent, banerji2019imaging, shrestha20LSA}, using various focal systems to obtain multiple focal length hence correcting the chromaticity \cite{wang2017broad, afridi2018electrically}, for imaging \cite{wang2018broadband, zhou2020flat} or image reconstruction through metalenses array \cite{zhu2020supercritical}, or implementation in quantum electrodynamics systems \cite{huang2019monolithic}. 
In this work, we propose a novel and effective approach for developing ultrathin all dielectric flat metalenses. The design involves generative adversarial networks (GAN) whereas, for the fabrication, a two photon direct laser writing (TP-DLW) process, upgraded to hyper resolution performance, is employed \cite{lio2020hyper}. The latter is achieved by exploiting a plasmonic nano-cavity metamaterial (Metal-Insulator-Metal-Insulator, MIMI) \cite{liocolor} that considerably improves the resolution of the TP-DLW process. The fabrication of 2D and 3D nanostructures with typical size of few tens of nanometers is thus enabled. The involved nano-cavity is an epsilon near zero ($\varepsilon_{NZ}$) metamaterial designed to support two resonances, centered at $\lambda=780nm$ and $\lambda= 390$ \cite{lio2020hyper}, more details in Method Section.
In the following, it is detailed how to engineer the design of ultra-thin flat metalenses, with thickness as low as 30-40nm, and their fabrication through an upgraded TP-DLW process able to pattern structures within less than a minute time. The latter has great significance for practical and industrial applications. Despite its ultra-thin thickness, the proposed metalenses show exciting features like apochromatic behavior, focal length of 1.14mm, numerical aperture of 0.87, FWHM of 0.9-1.6 $\mu m$ and depth of focus of 100-150 $\mu m$ in the visible range. These properties pave the way for the application of our metalenses in several technological fields like miniaturized photocameras for mobile phones, new concept microscopies, anticounterfeiting or telecommunication.
\section{Metalenses: numerical design and TP-DLW fabrication}
\begin{figure}[ht]
\begin{center}
	\includegraphics[scale=0.35]{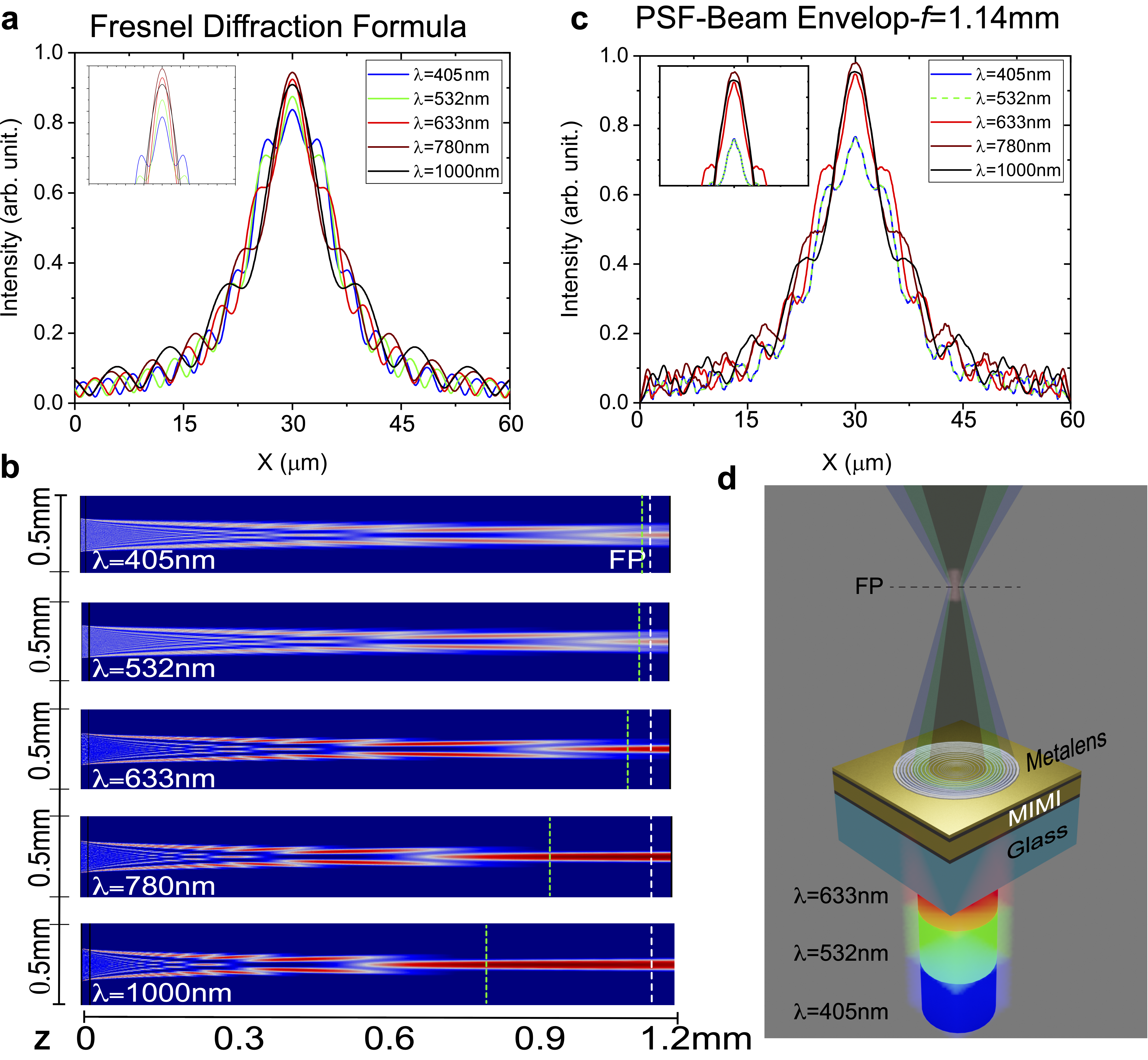}
	\caption{a) Beam envelope at the focal plane (common for five different wavelengths) calculated through the Fresnel diffraction formula by using the near field $E(x_0,y)$ distribution provided by the GAN algorithm. b) Electric field evaluated by using Finite Element Method (FEM) beam envelope method for five different wavelengths and c) Beam profile retrieved at the common focal plane (FP) by calculating the Point Spread Function (PSF); d) Schematic view of numerically designed achromatic metalens considering the hyper resolution achieved through the TP-DLW technique upgraded by a metal-insulator-metal-insulator (MIMI) nano-cavity.}
\label{1}
\end{center}
\end{figure}

The proposed metalenses reproduce the optical behavior of a classic lens by means of low-loss dielectric metasurfaces. However, their functionalities go beyond those of classical lenses by introducing a variety of optical modes enabling dispersive phase compensation and chromatic aberrations suppression \cite{park2019all}.  In order to efficiently design the metalenses, their geometrical parameters are retrieved through an inverse design method mediated by deep machine learning (DML). The latter exploits the "GLOnets" \cite{jiang2019global} and generative adversarial networks (GANs) to realize a target refractive index pattern. In detail, the algorithm iteratively modifies geometrical features of polymeric nanoridges like height $H$, width $W$, period $\Lambda$ and maximum transverse dimension $u$ (Figure S1a) until the metalens performs the desired optical function.
The range of sizes, within which these parameters can vary, are derived from the study detailing the hyper resolution TP-DLW technique \cite{lio2020hyper}. In particular, height and width for each element respectively vary from $5\pm 2$ to $40\pm 2 nm$ and from $200\pm 2$ to $500\pm 2 nm$. 
At each iteration, the developed GAN algorithm modifies the geometrical parameters to let the nanoridges design converge to desired focal length $f = (1 : 1.2 mm)$ and $u=(0 : 100 \mu m)$. At this point of the design, the spatial distribution of the nanoridges is 1D. The metalens with cylindrical symmetry is obtained by revolving the nanoridge pattern around its propagation axis. The GAN uses the actual geometrical parameters and the considered wavelength to calculate the produced electric near field $E(x_0,y)$ and compare it with the one that would be produced by an equivalent (same focus distance) classic lens. For each considered wavelength, namely $405nm, 532nm, 633nm, 780nm$ and $1000nm$, the calculated geometrical parameters that satisfy the comparison within a certain error are stored. Details about the adapted "GLOnets" algorithm are reported in the supplementary Figure S1b, some examples of the developed element arrangement are reported in Figure S2.  
By exploiting the following \textit{Fresnel diffraction formula}, the calculated near field $E(x_0,y)$ can be used to obtain the envelope $E(f,u)$ in far field of the wave propagating from the metalens. 
\begin{equation}
E(f,u)= \frac{1}{\sqrt{\lambda}}\int_{-\infty }^{\infty } E(x_0, y)e^{-i\pi y^2/(\lambda f)} e^{-i\pi uy/(\lambda f)}dy
\end{equation}
where $u/(\lambda f)$ is the spatial frequency, $E(x_0, y)$ and $E(f,u)$ are the electric field amplitude distributions at the exit plane $x=x_0$ and the focal plane $x=f$, respectively. From the above analysis, the metalens structure has the following retrieved parameters: total radius $u=100\mu m$, period $\Lambda=1\mu m$, height and width of the nanoridge element $H=30nm$ and $W=200nm$ respectively. In order to avoid hole diffraction, the inner center of the metalens is filled with lines. By using the previous formula with the retrieved geometrical parameters, the beam profile at the common focal length of 1.14 mm is evaluated for each considered wavelength; the results are reported in Figure \ref{1}a.

\begin{figure}[!h]
\begin{center}
	\includegraphics[scale=0.4]{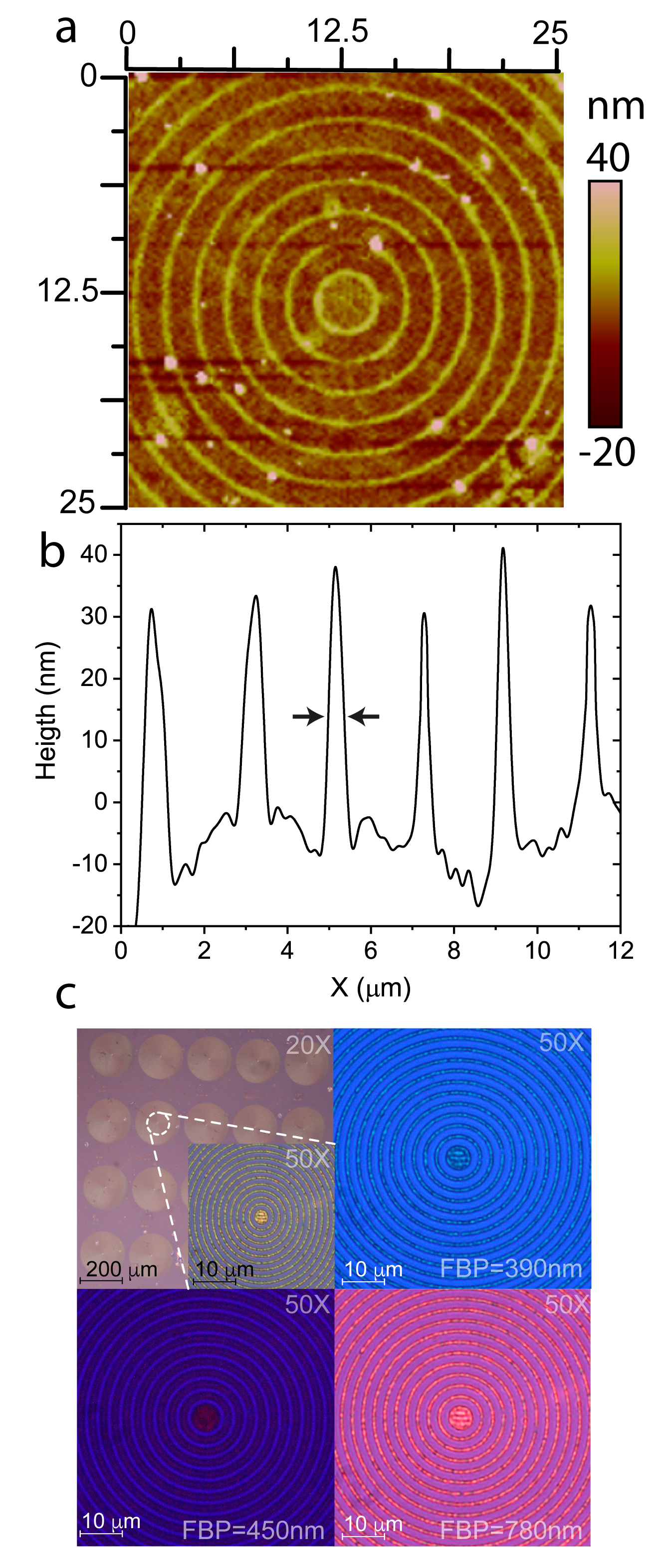}
	\caption{a) AFM morphological characterization of the metalenses with b) the related height and width profile of the circle constituting it. c) Optical image acquired with microscope with magnification 20x of the metalenses matrix, in the inset a zoom at 50x of one structure. Particular of one metalens at 50x observed using different band pass filter on the incoming light.}
\label{2}
\end{center}
\end{figure}

Figure \ref{1}b reports a further analysis conducted calculating the electric far field for each considered wavelength by means of a Finite Element Method (FEM) beam envelope model. The performance of an imaging system can also be quantified by calculating its point spread function (PSF). The amplitude PSF of a lens is defined as the transverse spatial variation of the amplitude of the image received at the focal plane (FP) when the lens is illuminated by a perfect plane source \cite{kino1996confocal} (Figure \ref{1}c). It is worth noting that the agreement between the results obtained using the different methods is quite impressive: each wavelength profile converges to the focal plane with a Full Width Half Maximum (FWHM) that is less than the diameter of the whole metalens, further confirming its achromatic behavior.  Thereafter, the depth of focus (DOF), or the distance from the focal point to the position where the intensity profile drops to half its maximum, is evaluated from the electric far field maps. This results close to $50 \mu m$ for shorter wavelength and more than $200 \mu m$ for longer ones, as indicated by dashed green lines drawn on Figure \ref{1}b. According with the performed numerical simulation and past studies \cite{khorasaninejad2017metalenses, decker2019imaging}, the designed metalenses present also a zero phase shift at the same focal length for all considered wavelengths, see Figure S3. This feature covers a fundamental role in the broadband achromatic lens projection. 
Once designed, the metalenses are fabricated exploiting an upgraded two-photon direct laser writing technique which adopts a metal-insulator-metal-insulator (MIMI) nano-cavity for the hyper resolution performance (Figure \ref{1}d).
In the last years, the TP-DLW technique has emerged as a new frontier for micro- and nano-fabrication often replacing electron beam and UV lithography. It is actually used to realize miniaturized efficient bulk optics \cite{gissibl2016two}. However, the present technology performance needs to be pushed to its limits to achieve the required patterning of all dielectric metasurfaces like ultra-thin diffractive optical elements \cite{chen2013reversible, chen2018broadband} or complex circular structures made by single pillars \cite{arbabi2020increasing}. A work from our group has recently demonstrated that a substantial improvement of the TP-DLW capabilities is obtained when the process is upgraded through the exploitation of a MIMI substrate. The introduction of the MIMI metamaterial in the TP-DLW process represents a ground-breaking technological leap: typical sizes of few tens and few hundreds of nanometers in height and slit width respectively \cite{lio2020hyper} are easily achievable and the fabrication of all dielectric ultra-thin nano-structures as flat metalenses is obtained in a fraction of the time required by other standard processes. The array of ($5X4$) metalenses considered in this work has been fabricated in less than five minutes thus confirming the fast fabrication procedure as extremely useful for mass production and application in end-user imaging devices\cite{zhu2020supercritical}.  
After writing and development steps, a morphological characterization on the fabricated metalenses has been performed using an Atomic Force Microscope (AFM), the acquired image are reported in Figure \ref{2}a. 
By showing a width $W\sim200 nm$ and a height of only $H\sim30 nm$, the AFM profile of each circle (Figure  \ref{2}b) confirms the ultra-thin characteristic of the proposed metalens  .

Finally, an optical microscopy analysis has been conducted highlighting the metalenses matrix and details of each one. Figure \ref{2}d shows an image acquired with a magnification 20x while the inset shows the zoom (magnification 50x) of a single structure. Successively, according with the peculiar feature of the nano-cavity \cite{liocolor},  a metalens has been observed including a band pass filter (BPF) centered at three different wavelengths, two close to the double $\varepsilon_{NZ}$ resonances and one far from them. The first test has been performed by using a filter centered at $\lambda_{BPF}=400 nm$. In this condition, the structure is very clear and it is also possible to distinguish horizontal lines in the inner circle of the metalens (Figure \ref{2}c, top right). These lines fill the inner center of the metalens to avoid hole diffraction. The same test has been repeated with a BPF centered at $\lambda_{BPF}=460 nm$, now the structure appears blur and the circles broad (Figure \ref{2}c, bottom left). Finally, the frame reported in Figure \ref{2}c bottom right has been collected by using a BPF centered at $\lambda_{BPF}\sim 780 nm$. As expected, the system appears again very clear and the features of each structure are highly defined. The high resolution at blue and infrared wavelengths is an effect related to the double resonances and a consequence of the zero light spreading property of the $\varepsilon_{NZ}$ MIMI metamaterial exploited during the fabrication process \cite{lio2020hyper}. Further detailed about the unusual optical behaviour of $\varepsilon_{NZ}$ metamaterials can be found in past studies\cite{fang2002imaging, casse2010super, zhao2011nanoscale, newman2013enhanced, shekhar2014hyperbolic}. 
\section{Experimental Results:\\ Focal Length of all-dielectric metalenses }
\begin{figure}[ht]
\begin{center}
	\includegraphics[scale=0.45]{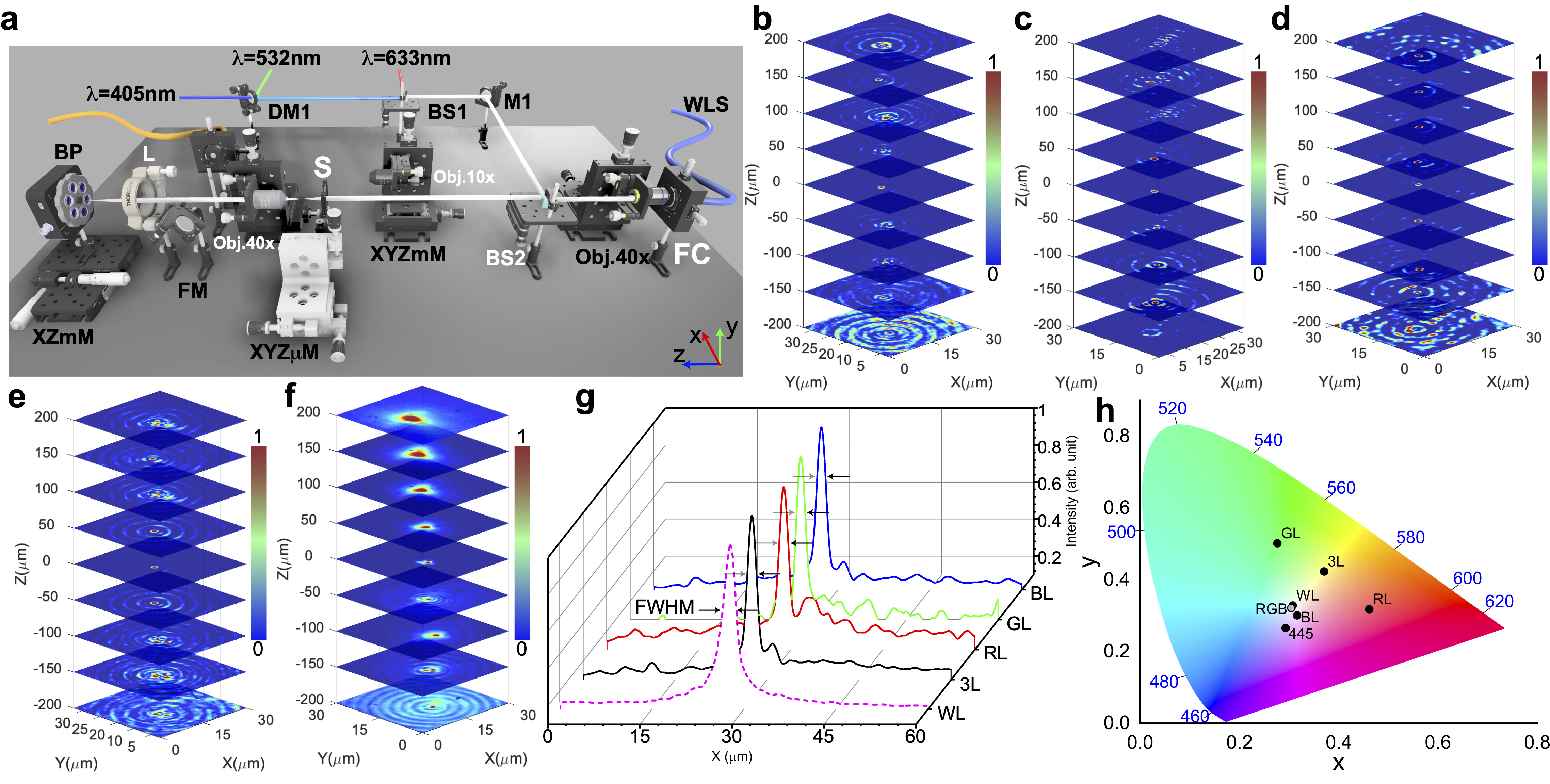}
	\caption{a) Schematic representation of the setup used to characterize the metalens. Stacks of the beam divergence from the focal length ($1.14 mm$) indicated as z=0 to $\pm 200 \mu m$ at b) $\lambda=405nm$ c)$\lambda=532nm$ d)$\lambda=633nm$ e) using the previous three lasers at the same time and f) using a Xenon lamp source. g) Full Width Half Maximum (FWHM) for all considered cases. It is $0.9 \mu m$ for each single laser (BL, GL, RL), $1.5 \mu m$ for the three lasers (3L) and $1.65 \mu m$ for the white lamp (WL). h) CIE chromaticity xy-coordinate plot for each laser wavelength, combination of them (3L) and white lamp (WL);  for laser with $\lambda=445 nm$  and the combination with the GL and RL producing RGB spot (grey circle).}
\label{3}
\end{center}
\end{figure} 
A complete optical characterization was performed on the fabricated metalenses exploiting the confocal home built setup shown in Figure \ref{3}a. More details are available in \textit{materials and methods}. 
The use of a beam profiler (BP) permits to collect, at the same time, the intensity of the signal, the beam profile,  high quality 2D pictures of the sample and a 3D reconstruction made by the intensity, as largely done in past studies \cite{she2018large, decker2019imaging, lee2020metasurfaces}. The focal length, the beam divergence and the point spread function (PSF) analysis has been performed using three lasers (3L), namely a blue one at $\lambda= 405nm$ (BL), a green one at $\lambda=532nm$ (GL) and red one at $\lambda=633$ (RL), and a white light source originating from a Xenon lamp (WL) with intensity of $50 \mu W/mm^{2}$. The response of the metalens at different propagation distances using each laser at one-by-one wavelength is reported in Figure \ref{3}b,c and d respectively. Then, the three lasers beams impinge on the sample at the same time to deduce the metalens behavior when enlightened by a white RGB coherent source (Figure \ref{3}e). Finally, the metalenses are studied under the effect of a broad range white light ($250 nm$ to $1100 nm$) as reported in Figure \ref{3}f. All the measurements validate the metalens broadband achromatic behavior with a focal length of 1.14 mm. The latter has been experimentally measured using the procedure reported in the supplementary Figure S4. 
In order to compare the numerical prediction and the experimental results, the full width half maximum (FWHM) at the focal plane ($f$) for each considered case has been evaluated and reported in Figure \ref{3}g. For each single laser wavelength (BL, GL and RL) the PSF presents a value of FWHM of only $0.9 \mu m$, for 3L the FWHM is $1.5 \mu m$ and for the WL it is $1.6 \mu m$. The experimental depth of focus (DOF) of the metalenses, which depends on the impinging wavelength, agrees with the numerical studies showing average values about $|100 -150| \mu m$ around the common focal distance for all wavelengths, see Figure S5 for more details. As described in the numerical section, the DOF is shorter in the visible spectrum than in the IR one.  Similar results have been demonstrated for grating-based metasurfaces \cite{bayati2020inverse}. The collected results lead to evaluate the numerical aperture (NA) of the proposed metalens to $NA\simeq nD/2f = 0.87$, details are reported in the supplementary information. Such high NA value is extremely useful for realizing very precise lens systems for imaging and confocal instruments.
Finally, in order to further confirm the achromatic behavior of the proposed metalens, a CIE chromaticity xy-coordinate graph has been evaluated for all the considered cases and reported in Figure \ref{3}h.  The use of a $\lambda=405 nm$ produces a blue/purple point and the three collinear lasers (3L) results shifted in a blur white on the tri-stimulus region (warm light)\cite{broadbent2004critical}. As expected, the white lamp is centered on the pure white point (cold light). In order to further confirm the RGB behavior, another test has been done using a laser with a different blue wavelength, $\lambda = 445nm$, obtaining this time a practically blue spot. Now, if this blue laser is combined with previous green (GL) and red (RL) lasers, a white spot comparable with the WL is obtained, reported in the gamut plot as the RGB grey circle. These results corroborate the broadband achromatic behavior of the proposed ultra-thin dielectric metalenses fabricated with a hyper resolute TP-DLW lithography process. 
The whole characterization confirms extremely exciting properties of the proposed metalens that can be exploited, due to their miniaturization, as optical components for micro-nano imaging, smart devices, in situ investigation as a local spectroscopic probe, color sensitivity and so on \cite{suwa2019min, huang2019monolithic, fan2019broadband, wei2020compact,  lee2020metasurfaces, malekovic2020soft,  kim2020fiber}. 
\section{Conclusions and Perspectives}
In this work, ultra-thin flat metalenses with vertical sizes in the order of few tens of nanometers have been reported. Design through GANs algorithms, fabrication by hyper-resolute TP-DLW process and optical characterization have bbeen demonstrated. The metalenses are able to work in a broad visible range focusing light at 1.14 mm focal lengtt, DOF of about $|100 -150| \mu m$ and high numerical aperture of 0.87. These metalenses present ground-breaking advantages as low-cost single step and time-saving fabrication, nanometric sizes, all dielectric composition. Moreover, they are engineered to be achromatic across the entire visible spectrum. The achieved results represent a significant advance in the state-of-the-art of research on metalenses, that are typically limited in their application by bandwidth and chromatic aberration.  These achromatic ultra-thin flat metalenses can find vast implementation across industry and scientific research, such as in miniaturized optical devices, advanced microscopy, local imaging and nano-lithography.
\section*{Materials and methods}
\textbf{Numerical model} The metalens design has been performed using numerical models implemented in Python and in COMSOL Multiphysics. The first step is using Python to code a generative adversarial networks (GANs) able to take in account all element variables. The developed GAN algorithms is based on "GLOnets" \cite{jiang2019global}. Then, by a training based on a deep machine learning (DML) process, the neural network returns the best parameters that satisfy the fixed conditions, in particular a common focal length at multiple wavelengths. 
Successively, the parameters retrieved using GAN code have been used to verify that the near field $E(x_0,y)$  produced by the metasurface satisfy the Fresnel diffraction formula ($E(f,u)$) and that the beam profile at the fixed focal length is maximum. Finally, in order to further validate the results in far-field numerical simulations, the beam envelop module of COMSOL Multiphysics has been used returning the electric far-field along the propagation direction. By placing a cut line at the focal length in the electric field map, the point spread function (PSF) of the propagated beam is retrieved. 
\\
\textbf{Metalenses fabrication} The nano-cavity has been deposited by DC sputtering on glass substrate with thickness of $\sim 100\mu m$) and it is constituted starting from the glass as follow: $30 nm$ Ag layer, $160 nm$ ZnO cavity, $30 nm$ Ag layer and $30 nm$ ZnO layer. The system is placed in front of the two-photon source. This lithography apparatus is used to define the metasurface lens pattern in a drop of photo-sensitive resist placed on top of the nano-cavity. The laser power used for the process is equal to $20 mW$ and the scan speed is $4000 \mu m$. After the writing process, the whole system (nano-cavity with cured and uncured photo-resin) is soaked in a bath of propylene glycol methyl ether acetate (PGMEA) for 15 minutes and then cleaned in a bath of isopropanol alcohol (IPA) for other 5 minutes. \cite{lio2020hyper} \\
\textbf{AFM characterization} AFM has been performed using a Zeiss LSM 780 with high resolute tips with precision of $\pm 2 nm$. Each scan has been collected at high resolution 1024x1024 px in order to reduce any background noise. 
\\
\textbf{Optical setup and characterization} A home-built setup composed of various optical elements has been used. A collimation line for  white light (WLS) that is coming from a Xenon lamp is realized with a 40X (Obj. 40X) objective and a fiber coupler (FC) mounted on a 3-axis stage (XYZmM) in order to collect the WLS by a fiber and send it collimated on the main confocal line. Then, on the same path it has been placed another 3-axis stage with an objective of 10X (Obj. 10X) used only to detect the metalenses on the sample. The objective is moved backward during the measurements in order to maintain the collimation of light impinging on the metalens. Beyond the metalens, there is an other objective 40X (Obj. 40X) that collects the signal from the sample and collimates it into the detector represented by a Thorlabs beam profiler (BP), Thorlabs BC106N-VIS spectral range from $350 nm$ to $1100 nm$,  mounted on a double-axis stage (ZXmM). Another lens (L) collects the signal from the objective and sent it on the BP ccd.
For acquire the spectrum needed to calculate the CIE chromaticity xy-coordinate, a flip mirror (FM) is placed before the collection lens aming to send the collimated signal on an iris that select only the focal spot. Then, the spectrum is acquired using a Flame Ocean Optics spectrometer. 
The metalens as moved forward and backward along Z in order to collect the beam divergence. Between the fiber collimator and the first objective it has been placed a beam splitter (BS2) that sends the light from lamp or from the laser (line) in the main path. The laser line is composed by a blue laser that sends the light through a dichroic mirror (DM1) that reflects the green and leads both lasers on a beam splitter (BS1) that collects the red laser and it sent all lasers on the second beam splitter (BS2).
\section*{Contributions}
G.E.L. conceived the idea, designed and numerically simulated the metalenses. G.E.L. improved the GLOnets code using a suitable GAN algorithms to design the metalenses. G.E.L., A.F. and T.R. fabricated and fully characterized the structures. G.E.L., A.F. and D.M.A. performed the optical measurements and analyzed the data. R.C, A.DL. and M.G. devised the experiments and supervised the work. G.E.L. and A.F. prepared and wrote the manuscript with input from all authors. 

\section*{Acknowledgements}
The authors thank the Infrastructure ``BeyondNano" (PONa3-00362) of CNR-Nanotec for the access to research instruments. They also thank the ``Area della Ricerca di Roma 2", Tor Vergata, for the access to the ICT Services (ARToV-CNR) for the use of the COMSOL Multiphysics Platform, Origin Lab, and Matlab.

\bibliography{bibliography} 
\newpage
\setcounter{figure}{0}
\renewcommand{\thefigure}{S\arabic{figure}}
\section*{Supplementary material}

\section*{Generative Adversarial Networks (GANs)}
\begin{figure}[ht]
\begin{center}
	\includegraphics[scale=0.2]{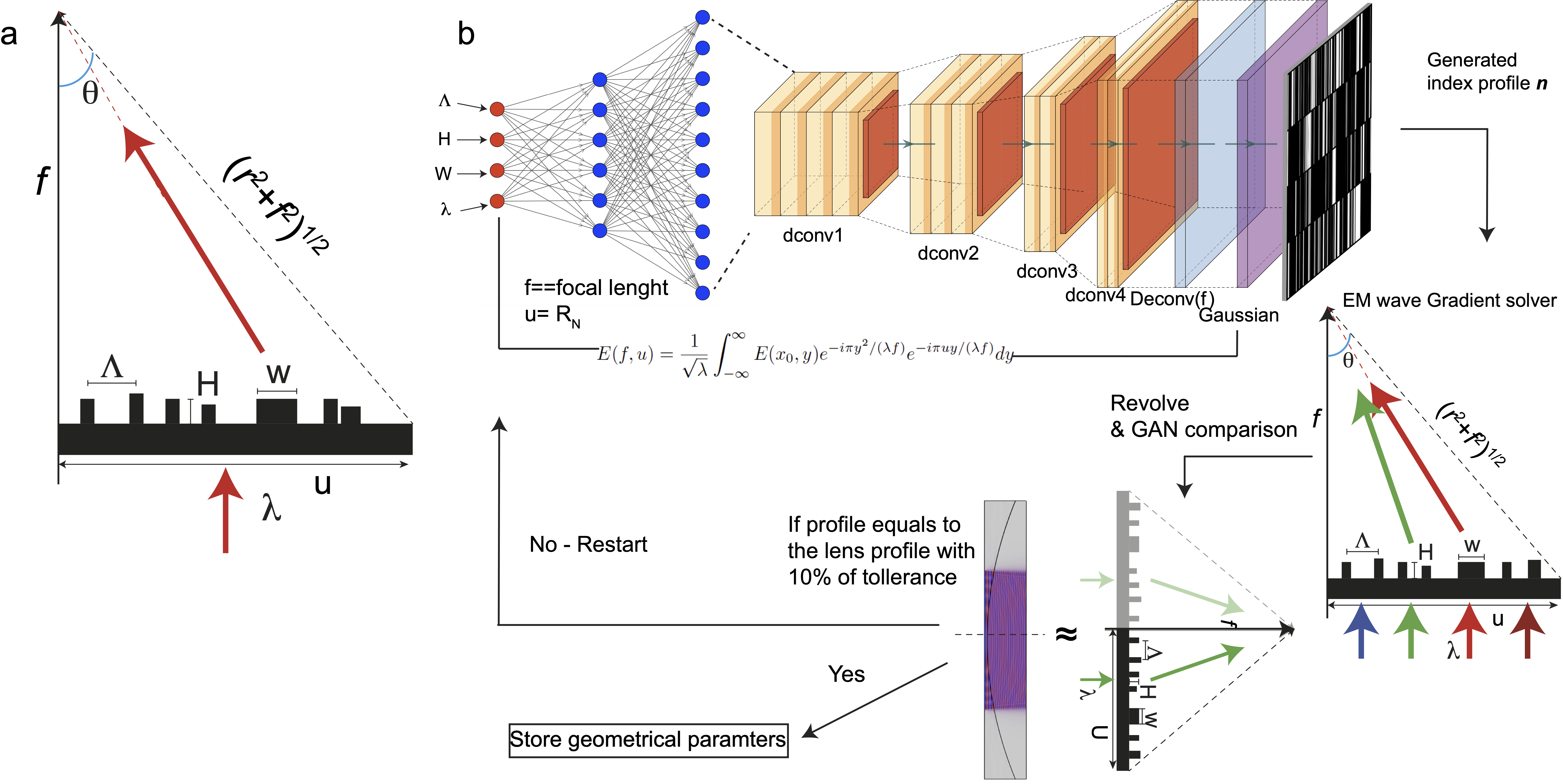}
	\caption{Global optimization based on a generative neural network (GAN) opportunely modified. a) Schematic view of  nano ridges that constitutes the metalens. b) Schematic representation of the conditional GANs code for metalens generation. The generator is built on fully connected layers (FC), deconvolution layers (dconv), and a Gaussian filter. An identity shortcut connection is also used by the electric near field equation. The input of GANs code are the device's geometrical parameters as height ($H$), width ($W$),  period ($\Lambda$) and finally wavelength ($\lambda$). During each iteration of training, a batch of devices is generated and for each device efficiency gradients (g) are calculated using forward electromagnetic simulations. These gradients are propagated through the network to update the weights of neurons ($W_n$). The GAN verifies that the near field E(x0; y) produced	by the metasurface satisfy the Fresnel diffraction formula	(E(f; u)). Then, the GAN algorithm compares the electrical near field of the actual geometrical parameters and the considered wavelength with the one of a classical lens having the same focal length.  Only the parameters that respect the fixed condition over the focal length and the maximum lens transversal dimension at multiple wavelengths are stored.}
\label{S1}
\end{center}
\end{figure}

For the metalens design we used a conditional Generative Adversarial Networks (GANs), based on GLOnets (Metanet code) \cite{jiang2019global},  that have wavelength, geometrical parameters, focal length and, if required, deflection angle as inputs. We designed and optimized an ensemble of polymer nano ridges that operate across a wide range of wavelengths in the UV-NIR range. This code builds on analysis of conditional GAN that can optimize only a single device in a training session. This means that  focal length and wavelength are fixed, while the other parameters are continuously changed from the input array values. The thickness of the nano ridges is varied in the range from $H\sim (5\pm2: 30\pm2) nm$ and the incident light is TM/TE-polarized. For each device, the nanoridge pattern period ($\Lambda$) is divided into segments $N = \Lambda/W$ where $W \sim (200\pm2: 500\pm2) nm$ represents the nanoridge width. For each segment, the refractive index of polymer or air is considered.
The nanoridge pattern deflects normally incident light to a direction meeting the focal point (Figure \ref{S1} a). The optimization scope is to maximize the deflection efficiency of the metalens given an operating wavelength ranging from 400 to 1000 nm and an outgoing angle ranging from $40^\circ$ to $80^\circ$. A schematic representation of our conditional GAN flow chart is presented in Figure \ref{S1}b. The input are the operating wavelength $\lambda$, the geometrical parameters $H$, $W$ and $\Lambda$, the desired focal length ($f$), the maximum transversal dimension ($u$) and the near field equation. The output is the refractive index profile n(u) of the device. The weights of the neurons are parameterized as $W_n$. The generator, conditioned on ($\lambda$, $H$, $W$ and $\Lambda$, $f$ , $u$) produces results which are compared with the near field profile of a classical lens. Due to the circular symmetry of a classical lens, the nanoridge pattern is then revolved around the propagation direction producing a circular pattern with the features designed by the mediated deep machine learning algorithm. This comparison is used to discard or store the obtained parameters. Finally, the parameters that allow a common focal length for all the considered wavelengths are used for the the numerical simulation using Finite Element Method models and the experimental fabrication/implementation.

\begin{figure}[ht]
\begin{center}
	\includegraphics[scale=0.35]{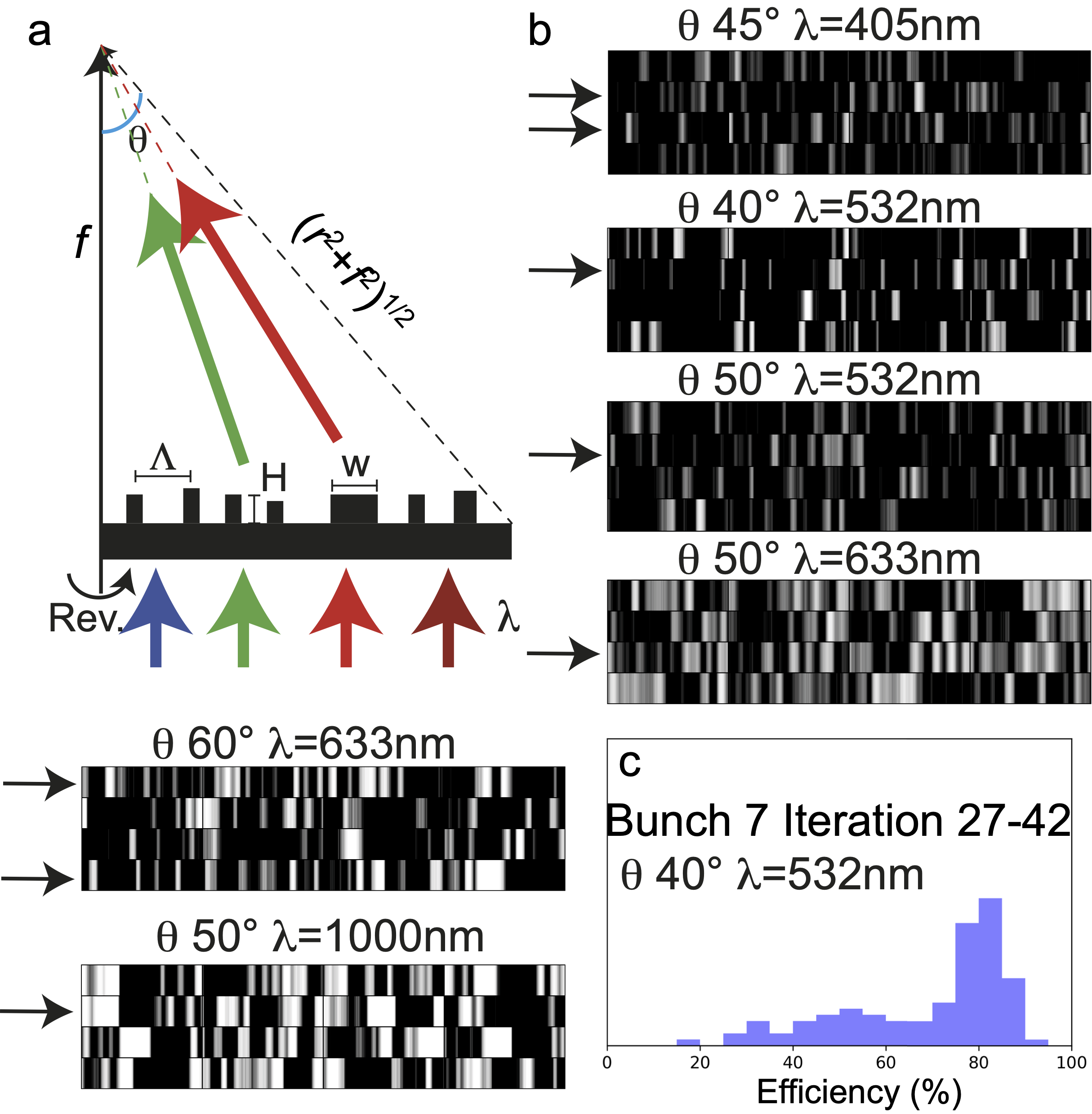}
	\caption{a) Sketch of the GAN algorithm working flow with the variable parameters. b) Examples of produced element disposition in function of wavelength and deflection angle, varying the height, width and period of the constitutive metalens element. c) Histogram efficiency used to chose the element disposition.}
\label{S2}
\end{center}
\end{figure}

Details of the GAN algorithms are reported in Figure \ref{S2}a depicting the working principle where each constitutive element is changed until the right values are reached. The algorithm has to satisfy an high value of efficiency for each considered wavelength for different deflection angles $\theta$. Figure \ref{S2}b reports the more efficient evaluated element disposition patterns for some cases $\lambda=405 nm$, $\lambda=532 nm$, $\lambda=633 nm$, $\lambda=1000nm$. For a bunch of iterations, the code returns an histogram with the efficiency reached for each disposition, as reported in Figure \ref{S2}c. Finally all the best data are stored and matched to obtain the element sizes that satisfy the initial condition as already discussed above. The enormous advantage of design algorithms coupled with the TP-DLW cad software is represented by the direct conversion of the produced design figure into TP-DLW writing file for a quick and immediate metalens fabrication. 
\section*{Phase shift at different wavelengths}
\begin{figure}[h]
\begin{center}
	\includegraphics[scale=0.4]{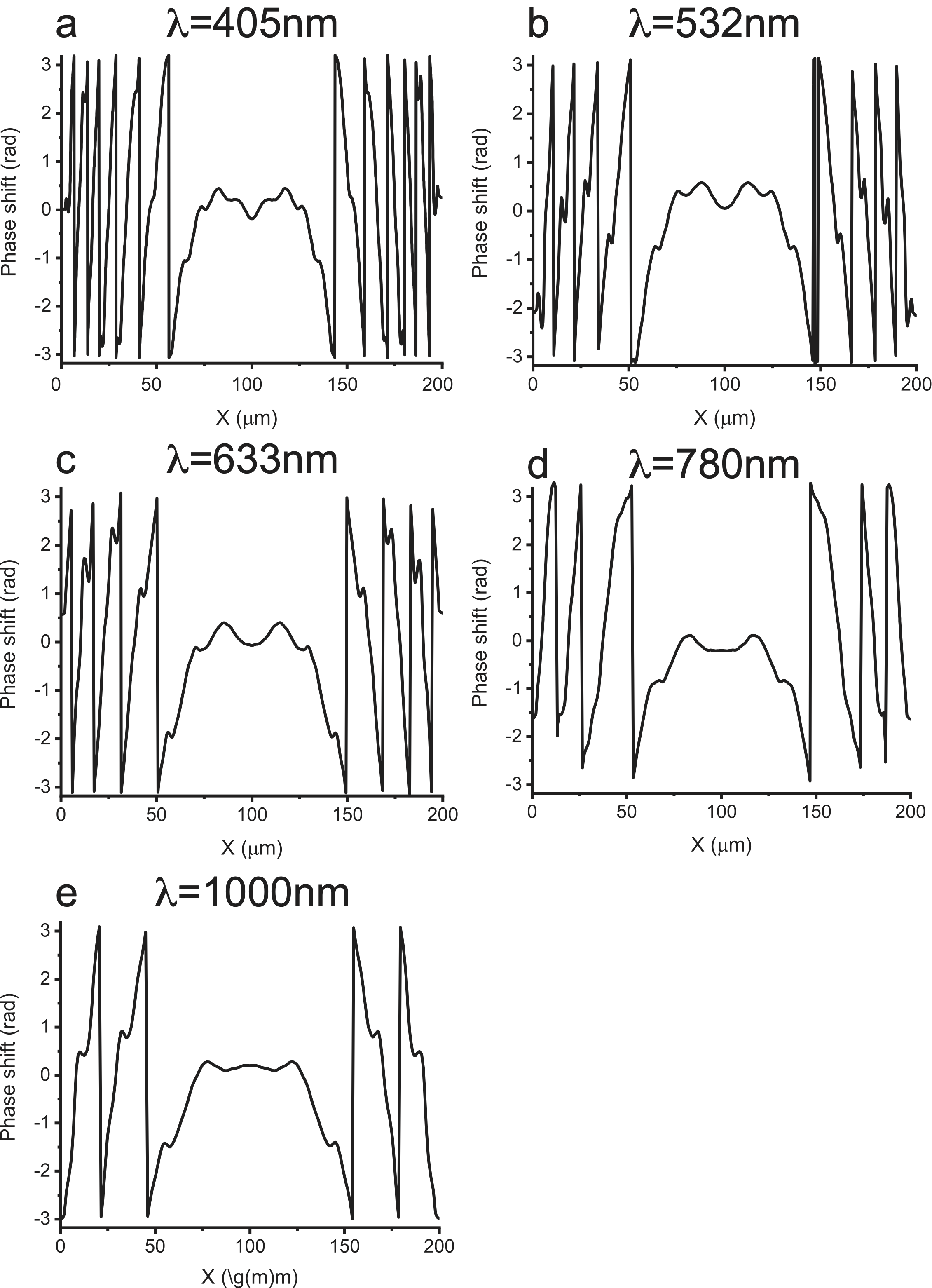}
	\caption{The phase shift analyzed for different wavelengths at the same focal length and along the lens diameter. The analyzed wavelengths are: a) $450 nm$, b) $532 nm$, c) $633 nm$, d) $780nm$ and finally e) $1000 nm$.}
\label{S3}
\end{center}
\end{figure}

The phase shift of the proposed metalenses has been numerically analyzed, see Figure \ref{S3}. It is close to 0 rad for all wavelengths at the focal length of $f=1.14mm$; similar results have been already presented for different achromatic metasurfaces\cite{khorasaninejad2017metalenses, decker2019imaging}. The phase is calculated as the argument of the electric field component along the propagation axis. The maximum transverse dimension of the nanoridges pattern "u", equal to $100 \mu m$, is the center of the metalens and the phase has been evaluated using a cut line at the calculate focal distance common for all wavelengths. The phase shift, according with past studies, results equal to $\phi(x)\sim 2\pi n_{eff}(x) t /\lambda$. Where x represents the dimension along the lens diameter with a maximum extension equal to $x=2*u$. When this equation is imposed equal to zero for each wavelength, the broadband achromatic lens design process starts \cite{aieta2015multiwavelength, genevet2017recent, chen2020flat}.
In order to clarify the connection between the already studied metalens systems and our GAN implementation, it is necessary to consider the phase equation written as a function of the radial coordinate $r$ resulting equal to $\phi(x) = -2\pi /\lambda_d (\sqrt(r^2-f^2)-f)$, with $r=\sqrt(x^2+y^2)$. Fixing $f$ and the maximum lens radius $r=u=(0:100)\mu m$, it is possible to run the GAN code until all conditions are respected and obtaining the element sizes ($H$,$W$, $\Lambda$) as already explained above. A comprehensive overview about the phase calculation and metalens design with different numerical tools are reported in \cite{chen2020flat}. 
\section*{Metalens focal length}
\begin{figure}[h]
\begin{center}
	\includegraphics[scale=0.5]{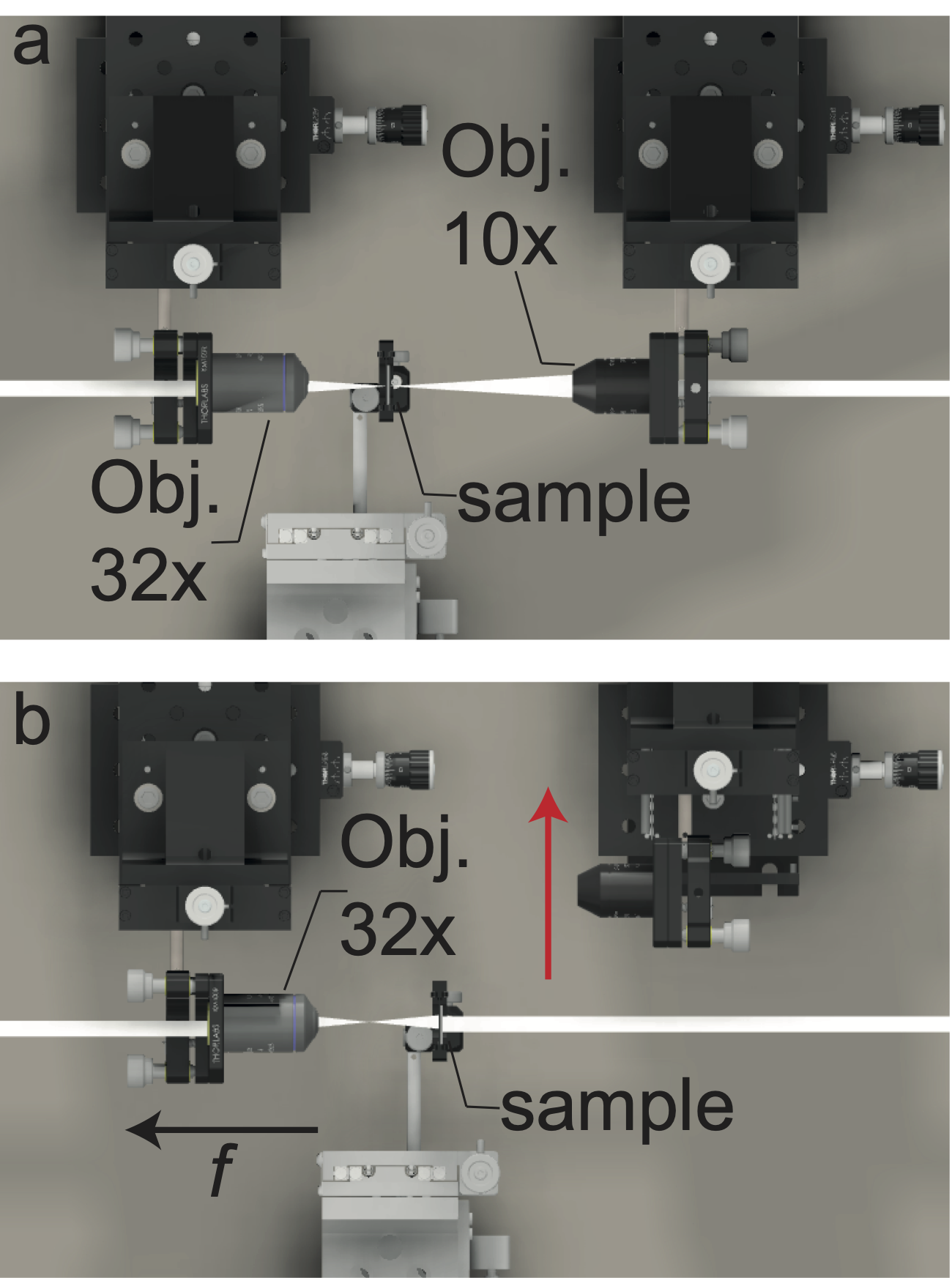}
	\caption{Sketch of the setup used to measure the focal length.  a) Initial confocal configuration. b) The metalens focal length measurements moving out the input objective and moving along z the output objective until the produced spot is again the smallest one.}
\label{S4}
\end{center}
\end{figure}

In this section, it is reported the procedure to measure the metalens focal length. A confocal homebuilt setup has been used allowing to remove the optical elements that are not necessary for the complete measurement. The first step consists to place the sample containing the metalens in the optical path of the confocal configuration (Figure \ref{S4}a). Then, once the focus is found on the metalens, the input objective (10X) is removed. Finally, the light impinges on the metalens and the output objective (32X) is moved along z until the new focal spot is established, Figure \ref{S4}b. The distance between the first configuration and the second one corresponds to the focal length, which in our case is $f=1.14mm$. 
\section*{Numerical Aperture (NA)}
The numerical aperture is evaluated using the following equation: 
\begin{equation}
NA=n sin \theta
\end{equation} 
In the condition of a low lens magnification the equation (1) is reduced to 
\begin{equation}
 {\text{NA}}_{\text{i}}=n\sin \theta =n\sin \left[\arctan \left({\frac {D}{2f}}\right)\right]\approx n{\frac {D}{2f}}
\end{equation} 
Where $D$ is the diameter of the lens $D=2*R_{N}$, $n$ is the refractive index of the metalenses $n\sim1.56$ and $f$ is the focal length. 
\section*{Depth of Focus (DOF)}
\begin{figure}[ht]
\begin{center}
	\includegraphics[scale=0.5]{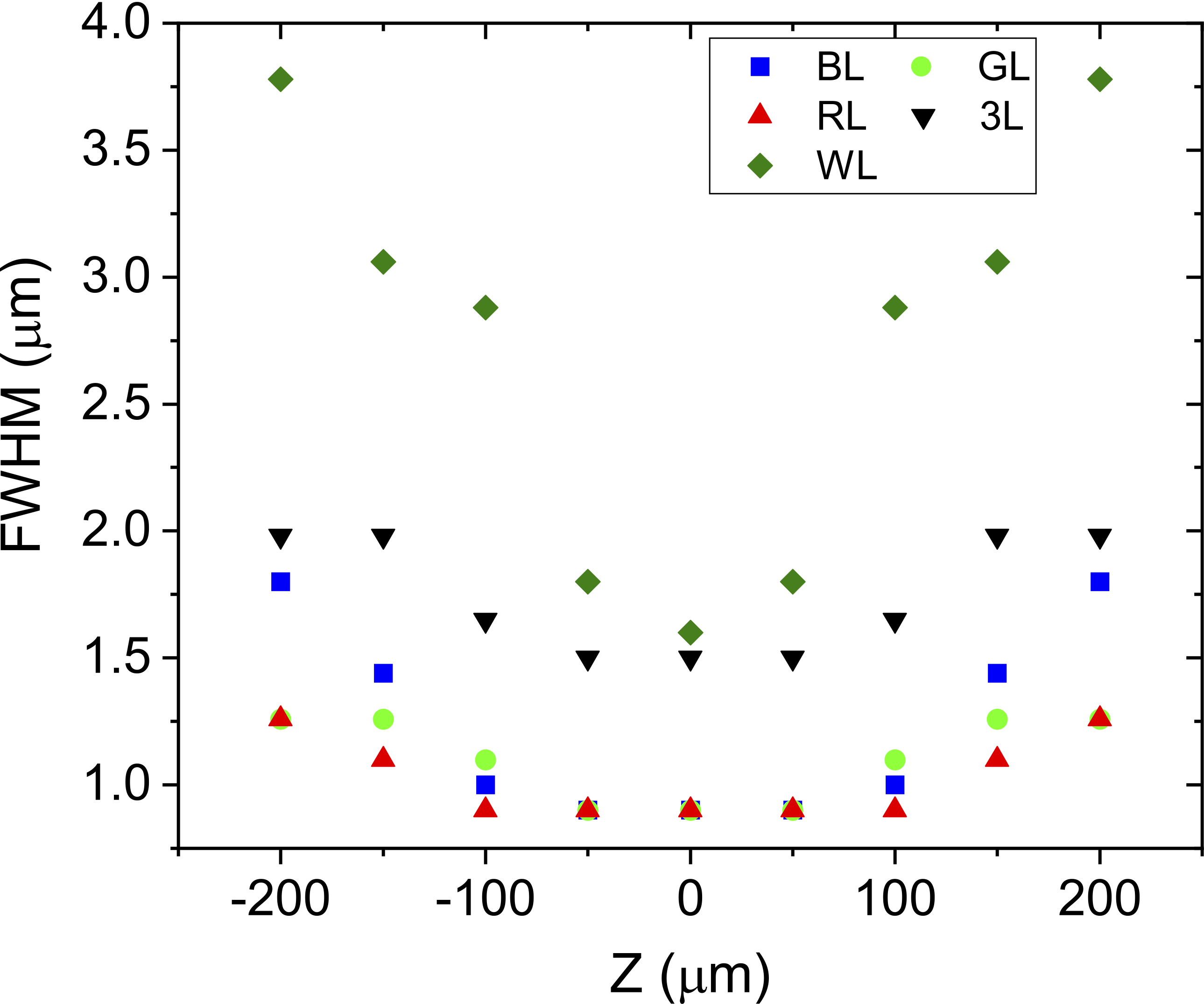}
	\caption{ Full Width Half Maximum (FWHM) of the produced spot is analyzed along the propagation direction $z$ in the range from $-200 \mu m$ to $200 \mu m$. It evidences the short DOF for the blu laser (blue squares), increasing a little bit for the green (green circles) and more for the red lasers (red up-triangles). The three lasers (3L) at the same time presents the more extended DOF that the previous cases (black down-triangles). The withe lamp (WL) due to the non coherence along the propagation presents a DOF reduced that the aspected one (dark green diamonds). }
\label{S5}
\end{center}
\end{figure}

The experimental DOF of the proposed metalens has been reported for each single light source and the combination of them, see Figure \ref{S5}. It is collected in the propagation range along $z$ from $-200 \mu m$ to $200 \mu m$. As predicted by the numerical simulations, the DOF is short in the blu-green spectral range, while it starts to increase when the wavelength is close to the red-IR spectral range. The combination of lasers (3L) presents a DOF extended as the red one. Instead, the withe light (WL) produces a depth of focus very short due to the incoherent nature of the propagated light and also because it is a continuous spectrum and the metalenses probably do not focus all wavelengths emitted by the Xenon lamp. 

\end{document}